\documentclass[twocolumn,showpacs,preprintnumbers,amsmath,amssymb,superscriptaddress]{revtex4}
\usepackage{placeins}
\usepackage{graphicx}
\usepackage{amsmath}
\usepackage{times}
\usepackage{color}

\begin{document}
\title{Reducing the Heterogeneity of Payoffs: an Effective Way to Promote Cooperation in Prisoner's Dilemma Game}

\author{Luo-Luo Jiang}
\affiliation{Department of Modern Physics, University of Science
and Technology of China, Hefei 230026 P. R. China}

\author{Ming Zhao}
\thanks{zhaom17@gmail.com} \affiliation{Department of Modern Physics, University of Science
and Technology of China, Hefei 230026 P. R. China}

\author{Han-Xin Yang}
\affiliation{Department of Modern Physics, University of Science
and Technology of China, Hefei 230026 P. R. China}
\author{Joseph Wakeling}
\affiliation{Department of Physics, University of Fribourg, Chemin
du Mus\'{e}e 3, CH-1700 Fribourg, Switzerland}

\author{Bing-Hong Wang}
\thanks{bhwang@ustc.edu.cn} \affiliation{Department of Modern Physics, University of
Science and Technology of China, Hefei 230026 P. R. China}
\affiliation{The Research Center for Complex System Science,
University of Shanghai for Science and Technology and Shanghai
Academy of System Science, Shanghai, 200093 P. R. China}

\author{Tao Zhou}
\affiliation{Department of Modern Physics, University of Science
and Technology of China, Hefei 230026 P. R. China}
\affiliation{Department of Physics, University of Fribourg, Chemin
du Mus\'{e}e 3, CH-1700 Fribourg, Switzerland}

\date{Received: date / Revised version: date}
\begin{abstract}
In this paper, the total payoff of each agent is regulated to
reduce the heterogeneity of the distribution of the total payoffs.
It is found there is an optimal regulation strength where the
fraction of cooperation is prominently promoted, too weak or too
strong of the strength will have little effects or result in the
disappearance of the cooperators. It is also found that most of
the cooperators are not distributed in isolation but form the
cooperator clusters, and to promote the cooperation the only way
is to enlarge the size of the cooperator clusters. Finally, we try
to explain the emergence of larger clusters and prove the
existence of the optimal regulation strength. Our works provide
insight into the understanding of the relations between the
distribution of payoffs and the cooperative behaviors.
\end{abstract}

\maketitle

\section{INTRODUCTION}
Cooperation is a widespread and important phenomenon in natural
and social systems, which is the foundation for the sustainable
development of creatures. However, individuals are instinctively
self-interested, which will drive them to cheat to obtain more
benefits rather than to cooperate. Therefore, understanding the
conditions for the emergence and promotion of cooperation is one
of the fundamental and central problems in biological, social, and
economic science
~\cite{inta1,inta2,inta3,inte1,inte2,inte3,inte4,inte5,inte6,inte7,inte8}.
Cooperative phenomenon can be well described by game theory. As
one of the representative games, prisoner's dilemma game (PDG)
seizes the characteristics of the conflict between the selfish
individuals and the collective interests. In PDG, when most of the
individuals take the cooperation strategy, the collective
interests is optimized, but as to an individual, if it cheats when
its opponents cooperate, it will profit much greater than it
cooperates and its opponents will profit little, even none. Thus,
more and more individuals will cheat, as a result the cooperation
will decrease. Ultimately, all the individuals will receive lower
payoffs than they take the cooperation strategy.

There are many mechanisms that can promote the cooperation of PDG,
such as repeated interaction~\cite{inta1}, spatial
extensions~\cite{inta4}, reciprocity~\cite{inta5}, and partly
randomly contacts~\cite{intd1}. Very recently, payoffs had also
been found playing an crucial role in promoting cooperation in
PDG~\cite{intc1,intf1,intf2}. Particularly, Perc found that
Gaussian-distributed payoff variations is more successful in
promoting cooperation than Levy distribution of
payoffs~\cite{intc1}, indicating heterogeneity of payoffs will do
harm to the cooperation amongst egoistic individuals. Then a
natural question has arisen: what is the effective way to regulate
the total payoff of each agent to optimize the cooperation? Here
in this paper, we try to give an answer to this question.

As we all know, taxation is of fundamental importance for every
country, it is not only a way to raise funds but also an effective
way to regulate the incomes of individuals to maintain the social
stability. In this paper, based on the idea borrowed from the tax
policy, we regulate the total payoff of each agent to decrease the
broadness of the distribution of payoffs. We will show that when
the payoffs are regulated to some extent, the fraction of
cooperation will be greatly promoted, however, if the regulation
is too strong, the cooperation will be suppressed again. We also
try hard to give a convincing explanation for the promotion of
cooperation.

\section{MODEL}
In the spatial PDG, agents located on a square lattice follow two
simple strategies: cooperation (C) or defect (D), described as the
form of vector:
\begin{equation}
\phi=\left(
  \begin{array}{c}
    1 \\
    0 \\
  \end{array}
\right)~{\rm or}~\left(
  \begin{array}{c}
    0 \\
    1 \\
  \end{array}\
\right).
\end{equation}
When a cooperator meets a cooperator, both of them get reward 1,
and when a defector meets a defector, they each get 0. And when a
cooperator meets a defector, it gets 0, but the defector receive
temptation $b$, $1<b<2$.  The above rule can be expressed by a
matrix:

\begin{equation}
\psi=\left(\begin{array}{cc}
      1 & 0 \\
      b & 0 \\
    \end{array}
  \right),
\end{equation}
which is called payoff matrix, and the parameter $b$ characterizes
the temptation to defection against cooperation. Each agent plays
PDG with its four neighbors. Therefore, the total payoff of the
player $i$ is the sum of payoffs after $i$ interacts with its 4
neighbors, which is written as:
\begin{equation}
P_i=\sum_{j\in\Lambda_{i}}\phi_i^{T} \psi \phi_j~,
\end{equation}
where $\Lambda_{i}$ denotes four neighbors of individual $i$. In
classical PDG, an agent updates its strategy according to the
following rule: the agent $i$ plays PDG with its neighbors, then
randomly selects a neighbor $j$, and adopts its strategy with
probability
\begin{equation}
G_{i \rightarrow j}=\frac{1}{1+{\rm exp}[(P_{i}-P_{j})/T]}~,
\end{equation}
where $T$ characterizes the stochastic noise. For $T=0$, the
individual always adopts the best strategy determinately, while
irrational changes are allowed for $T>0$. In numerical simulation,
noise level is often set as $T=0.1$ because a few irrational
behavior is common in real economic systems.

In our regulation scheme, we regulate the total payoffs:
\begin{equation}
W_{i}=P_{i}^{\alpha}~, \label{wfunction}
\end{equation}
where $\alpha>0$ is the regulation parameter. It is notable that
$\alpha=0$ is forbidden because it is meaningless for $P_{i}$. In
the case $\alpha=1$, the present model restores to classical PDG.
When $\alpha$ decreases from 1 to 0, the heterogeneity of the
payoffs distribution is also depressed, and when $\alpha$ goes to
0, the differences of total payoffs for different agents
disappear. In this paper, we replaced the total payoffs $P_i$ and
$P_j$ in equation (4) by the regulated payoffs $W_i$ and $W_j$,
and investigate how the cooperation will be affected by this
regulation.

\section{SIMULATION AND ANALYSIS}
n order to describe the evolution process of the game, we employ
the fraction of cooperations as an order parameter
\begin{equation}
\rho_{C}=\frac{1}
 {L^{2}}\sum_{i=1}^{{L}^{2}}\phi_{i}^{T}\left(
    \begin{array}{c}
      1 \\
      0 \\
    \end{array}
  \right).
\end{equation}
Based on a periodic boundary lattice with size of $100\times 100$,
an extensive Monte Carlo numerical simulation is performed with
random initial states. After the system reaches dynamic
equilibrium, $\rho_C$ is calculated and the final results are
obtained after the averaging of 10000 times.

\begin{figure}
\begin{center}
\includegraphics*[scale=0.80]{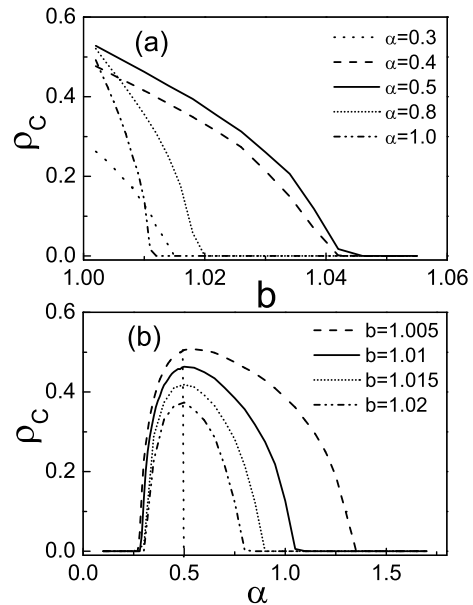}
\end{center}
\caption{Fraction of cooperation $\rho_{c}$ as a function of $b$
in (a) and $\alpha$ in (b). $b$ is fixed at $1.01$ in (a).
\label{cooperationration}}
\end{figure}

Figure~\ref{cooperationration} (a) shows the cooperation fraction
$\rho_{C}$ as a function of $b$ at different values of $\alpha$.
It displays that $\rho_{C}$ decreases monotonically with the
increasing of $b$, no matter what $\alpha$ is. Most interestingly,
the cooperation are greatly affected by the parameter $\alpha$ for
fixed $b$: in a large region of $\alpha$, $\rho_C$ will be
increased, indicating the reduction of heterogeneity of payoffs
will improve the cooperation. It worth noting that there is at
least one optimal value of $\alpha$ where $\rho_C$ takes its
maximum, larger or smaller $\alpha$ will cause the decreasing of
$\rho_C$. Thus, to quantify the effects of $\alpha$ on the
promotion of cooperation for different $b$, we present the
dependence of $\rho_{C}$ on $\alpha$ in
Fig.~\ref{cooperationration} (b). Clearly, in the $\alpha$ region
(0.37, 1.0), $\rho_C$ is larger than the case of classical PDG
($\alpha=1.0$), and at the point $\alpha=0.5$ $\rho_C$ reaches its
maximum, where the cooperation is promoted prominently. But when
$\alpha<0.2$ or $\alpha>1.4$, there will be no cooperator, which
means that both higher and lower value of $\alpha$ will jeopardize
cooperation. It is especially worth noting that our regulation
scheme is more powerful for larger temptation $b$, for example, at
$b=1.005$, for the classical PDG ($\alpha=1.0$), the fraction of
cooperation $\rho_C=0.3855$ and for the best case of our
regulation scheme ($\alpha=0.5$), $\rho_C=0.5060$, the increment
is $\Delta \rho_C=0.1205$; but for a larger temptation $b=1.020$,
the fraction of cooperation $\rho_C$ will increase from 0.0 to
0.3736, with the increment $\Delta \rho_C=0.3736$.

\begin{figure}
\begin{center}
\includegraphics*[scale=0.9]{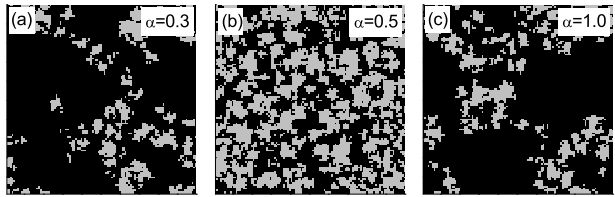}
\end{center}
\caption{For panels (a), (b), and (c), typical snapshots of the
distribution of cooperators (light gray) and defectors (black) on
a square $100\times 100$ lattice obtained for different value of
$\alpha$ by b=1.01. (a) $\alpha=0.3$, (b) $\alpha=0.5$, and (c)
$\alpha=1.0$. \label{payoff}}
\end{figure}

\begin{figure}
\begin{center}
\includegraphics*[scale=0.7]{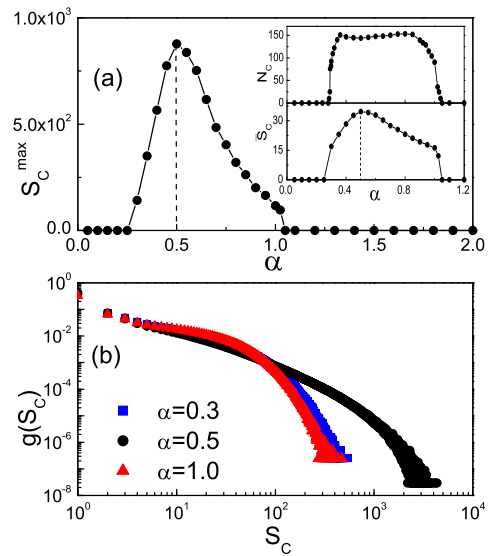}
\end{center}
\caption{(Color online) Panel (a) shows size of the largest
clusters formed by cooperators varying with $\alpha$, and the
inset shows the number of clusters $N_{C}$ (upper) and average
size of clusters formed by cooperators $\bar{S}_{C}$ (lower).
Clearly, $N_{C}$ keeps about $150$ from $\alpha=0.37$ to
$\alpha=0.85$, while $\bar{S}_{C}$ reaches its maximal value at
$\alpha=0.5$. Panel (b) displays distribution of cooperators'
cluster size for different values of $\alpha$. b=1.01 in both
panels (a) and (b). \label{distribution}}
\end{figure}

To explain why the cooperation are so prominent at $\alpha=0.5$,
we give the snapshots of the distribution of cooperation when
$\alpha$ takes three typical values: 0.3, 0.5 and 1.0, which are
shown in Fig~\ref{payoff}. Clearly, most of the cooperators are
not distributed in isolation but form some clusters. When the
payoffs are not regulated ($\alpha=1.0$) or regulated too much
($\alpha=0.3$), there are only a few cooperator clusters in the
system, but when $\alpha=0.5$, there will emerge so many
cooperator clusters that the cooperation are remarkably promoted.
We also count the number of cooperator clusters, which has no
notable changing for a large region of $\alpha$, but the maximal
and average size of the cooperator clusters will be greatly
increased in some region of $\alpha$, which is shown in
Fig~\ref{distribution}(a) and the inset. Moreover, the
distribution of cooperator clusters at the three values of
$\alpha$ are also plotted in Fig~\ref{distribution}(b). It is
clear, when $\alpha=0.5$, there are much more large clusters
($S_{C}>86$) than when $\alpha$ is 0.3 or 1.0, and the small size
clusters ($4<S_{C}<80$) are less than the two cases.

From the above simulation results we can assert, when $\alpha$
increases from 0.0, there is no cooperator in the system until
$\alpha$ reaches some threshold, then the cooperators emerge, with
$\alpha$'s increasing, more and more agents becomes cooperators
and the cooperator clusters become larger and larger, when
$\alpha$ reaches 0.5, the number of cooperators, the maximal and
average size of cooperator clusters all reach their maximums, and
further increase $\alpha$, where the payoffs are not regulated
much, the cooperations are repressed again, more and more agents
prefer to cheat, the cooperation decrease again, till disappear.
It seems that the emergence of larger cooperator clusters cause
the promotion of cooperation since when the fraction of
cooperation is promoted the number of cooperator clusters will not
change but the size of them will be greatly enlarged, as shown in
insets of Fig~\ref{distribution}(a). Our assertion is consistent
with the previous researches that cooperators survive by forming
compact clusters. And the cooperative agents along boundary
resisting against defectors can be enhanced by, for example,
heterogeneous structure~\cite{resa1}, attractiveness of the
neighbors~\cite{resa2, resa3}, and stochastic
interactions~\cite{resa4}.

Here we try to explain how the cooperator clusters are formed and
why the average cluster size reaches maximum at $\alpha=0.5$. In
our simulation, the initial state, cheat or cooperate, of each
agent is assigned randomly, from the statistical theory we know
that at the very beginning there are already some cooperator
clusters, to enlarge the already exist clusters, the defectors
along the boundary of the cooperator clusters should be more
inclined to change their strategies to cooperate, and the
probability is decided by two aspects, one is the number of
cooperators of their 4 neighbors and the other is the probability
$G_{i \rightarrow j}=1/\{1+{\rm exp}[(W_{i}-W_{j})/T]\}$. Along
the boundary of the cooperator clusters, the defector must have
neighbors who are cooperators, so their total payoffs $P_D$ are
not less than $b$, and for the cooperators, they must have at
least one neighbor who is also cooperator, which makes their total
payoffs $P_C$ take values 1, 2 or 3 but 0, as a result, $P_D$ may
larger or smaller than $P_C$. When $P_D>P_C$, with $\alpha$'s
decreasing from 1, $P_D^{\alpha}-P_C^{\alpha} > 0$ becomes smaller
and smaller, correspondingly, $G_{D \rightarrow C}$ becomes larger
and larger. But when $P_D<P_C$, with $\alpha$'s decreasing from 1,
$P_D^{\alpha}-P_C^{\alpha} < 0$ becomes larger and larger,
correspondingly, $G_{D \rightarrow C}$ becomes smaller and
smaller. As a consequence, when $P_D>P_C$, smaller $\alpha$ is
better for the formation of cooperator clusters but when
$P_D<P_C$, larger $\alpha$ is better, thus, there must be a right
value of $\alpha$, which is optimal for the formation of the
cooperator clusters, that is the reason why the fraction of
cooperator is optimized at $\alpha=0.5$ in our simulation.

\begin{figure}
\begin{center}
\includegraphics*[scale=0.7]{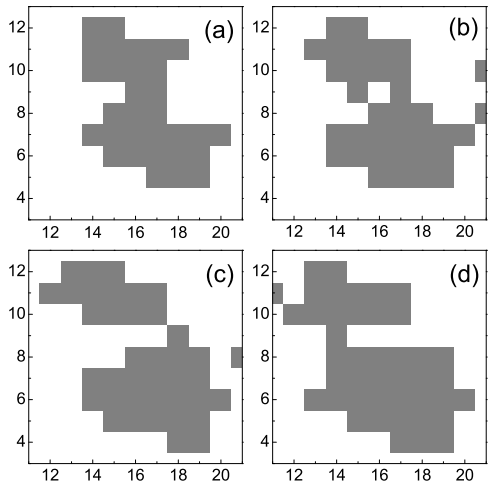}
\end{center}
\caption{Four successive evolvements of the cooperator clusters.
From (b) to (c), a cluster is divided into two clusters and from
(c) to (d), two clusters form a larger new one. Here, the area of
$10\times 10$ is the part of $100\times 100$ in the system, and
the light gray sites present cooperative agents. \label{Fig4}}
\end{figure}

From the above analysis we can also concluded that the more
cooperators of a defector's neighbors are, the easier for it to
become a cooperator. The reverse situation can also be proved that
if most of the neighbors are defectors for a cooperator, the
easier for it to becomes a defector. By far, we can give a picture
of the evolution of the cooperator clusters: there must be broad
boundary where the defectors around are changed to cooperators and
acute boundary where the cooperators become defectors, which can
be seen in Fig.~\ref{Fig4}. A larger cluster can be divided into
some smaller clusters when it is cut by defectors
(Fig.~\ref{Fig4}, from (b) to (c)), that is the reason why the
clusters could not keep growing. Certainly, several clusters can
also form a large one (Fig.~\ref{Fig4}, from (c) to (d)). So the
clusters evolve endlessly in a system.

\section{Conclusion}

In conclusion, we regulate the total payoffs of each agent to
narrow down the differences between agents in the spatial
prisoner's dilemma game, and find that there is an optimal
regulation strength where the cooperation is greatly promoted,
especially for larger temptation. Too strong of the regulation
will depress the cooperation, even cause the disappearance of the
cooperators. We reassure that it is the larger size not the number
of cooperator clusters that promote the cooperation. We also prove
the existence of the optimal regulation strength and explain the
formation of larger cooperator clusters. Our results provide
quantitative analysis of payoffs' effects on cooperation in
spatial prisoner's dilemma game. Compared with other factors of
maintaining cooperation such as spatial extensions, reciprocity,
and punishment, reducing the heterogeneity of payoffs is more
realistic for it seizes individual fitness by game payoffs.
Regarding economic process, our findings suggest that moderate tax
policy can plays an important role in maintaining cooperation
among social individuals.

\acknowledgments{The authors would like to thank Dr. Wen-Xu Wang
for his assistances in preparing this paper. This work was
partially supported by NSFC(Grant Nos. 10805045, 10635040,
60744003, 70871082), and the Specialized Research Fund for the
Doctoral Program of Higher Education of China (SRFD No.
20070420734).}

\bibliographystyle{h-physrev3}

\end{document}